\documentclass[iop,revtex4]{emulateapj}
\usepackage{subfigure}
\usepackage{amsmath}
\usepackage{amssymb}
\usepackage{bm}
\usepackage{url}
\usepackage{physics}
\defcitealias{MD14}{MD14}

\newcommand\myprime{\raise0.6ex\hbox{$\scriptstyle\prime$}}
\newcommand{\mdeg}{\ensuremath{^{\circ}}}

\newcommand{\rlc}{\ensuremath{\rm R_{\rm LC}}}

\newcommand{\dTheta}{\ensuremath{\Delta\Theta}}

\newcommand{\ppp}{\ensuremath{\rm P_{3}}}
\newcommand{\pppp}{\ensuremath{\rm P_{4}}}
\newcommand{\pppobs}{\ensuremath{\rm P_{3}^{\rm obs}}}

\newcommand{\nsp}{\ensuremath{N_{\rm sp}}}
\newcommand{\phizero}{\ensuremath{\phi_{0}}}
\newcommand{\phia}{\ensuremath{\phi_{1,\nu_1}}}
\newcommand{\phib}{\ensuremath{\phi_{1,\nu_2}}}
\newcommand{\sgn}{\mathop{\mathrm{sgn}}}
\newcommand{\dtobs}{\ensuremath{\Delta t_{\rm obs}}}
\newcommand{\s}[1]{\ensuremath{S_{#1}}}
\shorttitle{Expected imprints of the carousel}
\shortauthors{Maan, Y.}
\begin{document}
\title{Expected imprints of the carousel in multi-frequency pulsar observations\\ and new evidence for multi-altitude emission}
\author{Yogesh Maan}
\email{maan@astron.nl}
\affil{ASTRON, Netherlands Institute for Radio Astronomy, Oude Hoogeveensedijk 4, 7991 PD, Dwingeloo, The Netherlands}

\begin{abstract}
Subpulse modulation has been regarded as one of the most
insightful and intriguing aspects of pulsar radio emission.
This phenomenon is generally explained by the presence of a carousel of sparks
in the polar acceleration gap region that rotates around the magnetic axis
due to the E$\times$B drift. While there have been extensive single pulse
studies, geometric signatures of the underlying carousel, or lack thereof, in
simultaneous multi-frequency observations have remained largely unexplored.
This work presents a theoretical account of such expected signatures, particularly
that of a geometry induced phase-offset in subpulse modulation, including
various formulae that can be readily applied to observations.
A notable result is a method to resolve aliasing in the measured subpulse
modulation period without relying on knowledge of the viewing
geometry parameters. It is also shown in detail that the geometry induced
phase-offset enables critical tests of various observed phenomena as well
as proposed hypotheses, e.g., multi-altitude emission, magnetic
field twisting, pseudo-nulls, etc., in addition to that of the carousel model
itself. Finally, a detailed analysis of a 327\,MHz pulse-sequence of PSR~B1237+25
is presented as a case study to test the single-frequency multi-altitude
emission scenario. The analysis provides a firm evidence of inner and outer
conal components of this pulsar to have originated from the same carousel of
subbeams and emitted at different heights.
\end{abstract}

\keywords{Pulsars: general, pulsars: individual (B1237+25),
radiation mechanisms: non-thermal, stars: neutron}
\section{Introduction} \label{sec_intro}
A considerable fraction of pulsars exhibit systematic variation in
position and intensity of single pulse components when viewed from
one pulse to other. The single pulse components are called subpulses,
and a systematic variation in their position make them appear to drift
in pulsar's rotation phase. This fascinating phenomenon is known as
subpulse drifting \citep{DC68}, and has been regarded as one of the most
insightful aspects of the pulsar radio emission. The phenomenon is
generally understood in the context of the widely adopted ``carousel
model'' \citep{RS75}. This model attributes subpulse drifting to
the presence of a carousel of regularly spaced `sparks' circulating
around the magnetic axis in the polar cap region of the star because
of the E$\times$B drift. The sparks give rise to plasma beams streaming
upward from the surface along the magnetic field lines. At different
altitudes, these ``subbeams'' emit radiations in different radio
frequency ranges. A single carousel could give rise to subpulse
modulation in a maximum of two pulse components. To explain modulation
in more than two components observed for
several pulsars, \citet{GS00} proposed a maximal packing of
the sparks and effectively presence of more than one carousel on the
polar cap. More recently, it has been suggested that the E$\times$B drift
of the sparks is around the point of maximum potential at the polar cap
\citep{JvL12,Szary17} which could help in explaining some other
observed modulation varieties such as bi-drifting.
\par
The subpulse modulation properties are important observables that enable
probes of the plasma configurations in the magnetosphere as well as help in
examining various physical models. For example, coherent subpulse modulations
in single
pulse sequences of a few pulsars have been used to map the accessible parts
of the subbeam configurations \citep[e.g.,][]{DR99,DR01,MR08,MD14}. The
expected subpulse modulation phase, for a given pulse longitude, depends
on the viewing geometry as well as the number of sparks in the carousel.
This relationship enables constraining the circulation period of the
carousel for a known emission geometry. The modulation phase envelope
across the pulse-profile has also been used to scrutinize the carousel model
or its subsequent variants \citep[e.g.,][]{Edwards03,MD14}.
\par
Several of the observed pulsar properties depend on the altitude where the
radio emission originates. For example, the pulsar radio emission beam width,
and hence the observed pulse-width becomes larger at low frequencies due
to the dipolar flaring of the field lines \citep[][]{Cordes78}. The position
angle of the linearly polarized intensity also depends on the magnetic
configuration at the emission site. Depending on the emission altitude,
the average total intensity and polarization position angle profiles also
show effects of aberration and retardation \citep[][]{BCW91}. However,
the subpulse modulation
phase is primarily decided by the circulation of the carousel at the polar cap
and how the plasma sub-beams propagate to the emission sites. The foot-points
of the magnetic field lines that give rise to the observed emission at
different altitudes differ slightly in their magnetic colatitudes at the
polar cap. Hence, the observed emission at different altitudes samples
slightly different parts of the seed spark pattern at the polar cap at
its different rotation phases. This work presents the implications of
this aspect in multi-frequency and wide-band observations of subpulse
modulations in pulsars\footnote{During manuscript preparation
\citep[cf.][]{Maan18}, a similar description was presented by \citet{Bilous18}
for PSR~B0943+10. Our independent formulation presented in Section~2.1
introduces a minor improvement in the expression given by \citet{Bilous18},
and also demonstrates how the phase-offset depends on other pulsar quantities.}.
The preliminary results of this work were presented in the conference
``Pulsar Astrophysics $-$ The Next 50 Years'' \citep{Maan18}. 
\par
Section~2 provides a detailed physical description and formulation of the
frequency-dependent subpulse modulation phase-offsets that are expected
to be observed for a given viewing geometry and demonstrates the effect
using simulations. This section also presents methods to utilize the above
expected phase-offsets to determine several physical parameters related
to the carousel. The frequency-dependent phase-offsets also enable tests
of several observed or proposed phenomena that are often discussed within
the scope of the carousel model. Some of these tests are briefly described
in Section~3.
Section~4 presents a practical application of this aspect using a sensitive
single pulse sequence of PSR~B1237+25 and provides firm
evidence for single frequency multi-altitude emission in this pulsar. The
overall conclusions are presented in Section~5.
\begin{figure*}
\centering
\subfigure[]{\includegraphics[width=0.47\textwidth,angle=0]{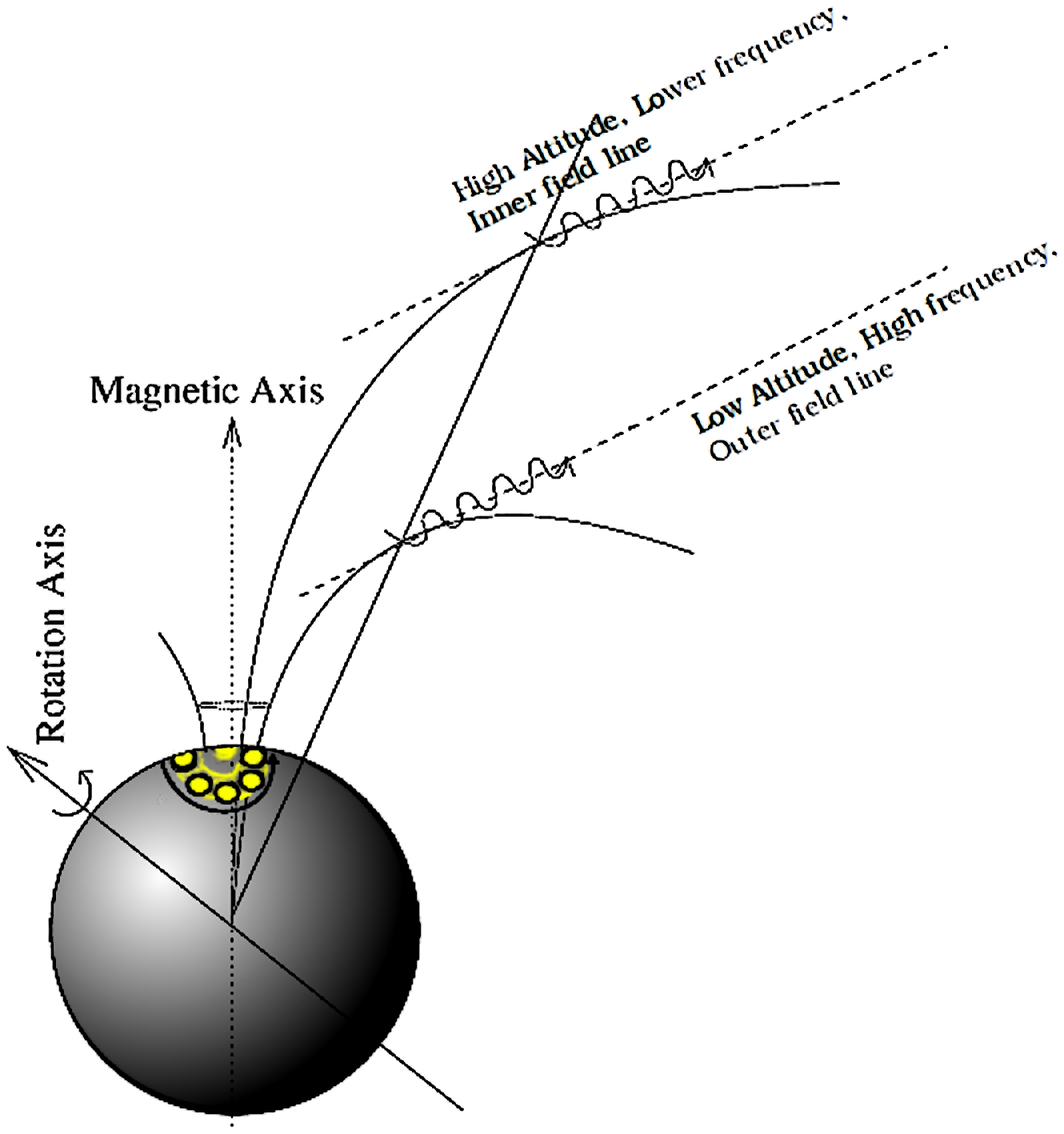}
\label{fig_rfm1}}
\hspace*{2mm}
\subfigure[]{\includegraphics[width=0.47\textwidth,angle=0]{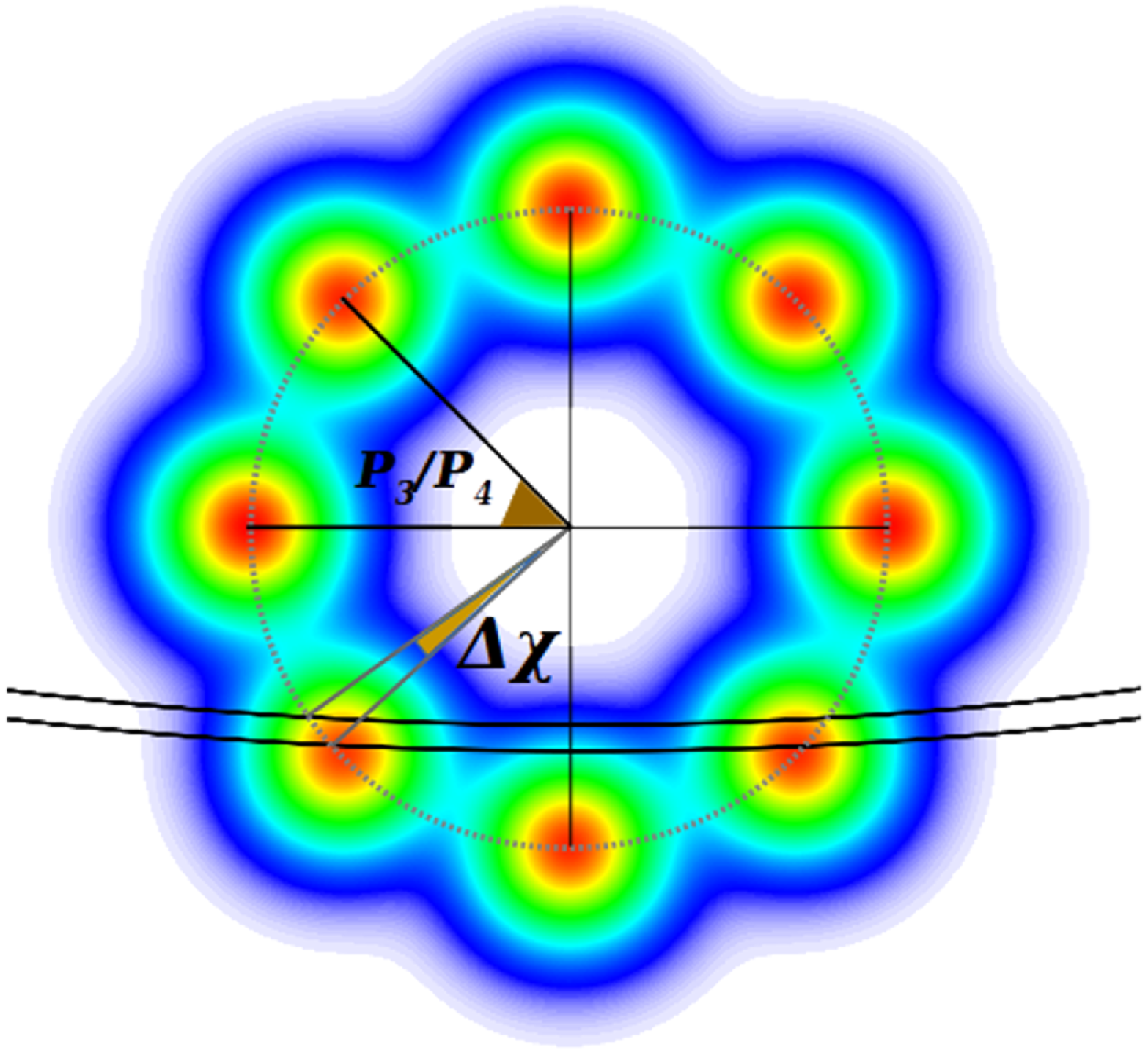}
\label{fig_rfm2}}
 \caption{\textsl{Left:} A demonstration of radius-to-frequency mapping
of radio emission from pulsars. The figure is not up-to-scale, primarily
to show that the high frequency emission originates at a lower
altitude and from slightly outer field lines when compared to the low
frequency emission.
\textsl{Right:} A simulated carousel of sparks and sightline
traverses corresponding to 240 and 610\,MHz traced back to the pulsar surface.
The inner (poleward) traverse correspond to the lower frequency. It is clearly
evident that the two frequencies would sample the carousel at its different
rotation phases. The instantaneous offset in magnetic azimuth is denoted by
$\Delta\chi$. The angular separation of consecutive sparks is proportional
to the ratio of the subpulse modulation and the carousel circulation periods
(\ppp/\pppp).}
\label{fig_rfm}
\end{figure*}

\section{Frequency-dependent subpulse modulation phase-offset: formulation and demonstration}
Radius-to-frequency mapping \citep[hereafter RFM;][]{Cordes78} suggests that
the radio signals at different frequencies originate at different
altitudes from the pulsar's surface, with lower frequency signals being
emitted farther away.
Furthermore, for a given viewing geometry, the observed emission originates
from a set of field lines that are parallel to the observer's line-of-sight
at the emission altitude (see Figure~\ref{fig_rfm1}).
This aspect, along with the dipolar configuration
of the magnetic field lines and RFM, implies that the observed emission at
different frequencies originates from slightly different sets of fields lines,
with the lower frequency emission originating from (magnetic) poleward field
lines \citep[see, for example,][]{DR01,ES03,Maan13}. Hence, the observed
radio emission in different frequency ranges sample slightly
different parts of the polar-cap and active plasma processes therein, e.g.,
the carousel of sparks (see Figure~\ref{fig_rfm1}).
Since the observed subpulse modulation results
from a set of emission sub-beams that stem from, and \emph{co-rotate} with,
a carousel of sparks on the polar-cap, observations at different frequencies
would sample the carousel at its different rotation phases.
\par
Figure~\ref{fig_rfm2} presents a simulated carousel of sparks on the
polar cap along with the \emph{footpoint}-traverses of field lines responsible
for emission at two different frequencies for a given viewing geometry.
The assumed frequencies are 240 and 610\,MHz, and the corresponding
emission altitudes are estimated to be about 710 and 560\,km,
respectively, using the empirical
formula given by \citet[][see their equation 3]{KG03}.
The viewing geometry parameters
(magnetic inclination angle, $\alpha$=50\mdeg, and impact angle
$\beta$=2.6\mdeg) and the rotation parameters (period, $P$=1.28\,s,
and period-derivative, $\dot{P}$=6.8e$-15$) are assumed to be those known
for the bright radio pulsar PSR~B0834+06. From the two emission heights, the
magnetic field lines corresponding to the sightline traverse are traced back
up to the pulsar surface to determine the corresponding footpoint traverse.
As expected, the lower frequency emission samples poleward part of the polar
cap. This fact has been exploited to map larger fraction of the underlying
emission patterns utilizing the circulation of the carousel
\citep[e.g.,][]{DR01,AD01,Maan13}. The peaks of the pulse components
correspond to field lines which have their footpoints on the ``circumference''
of the carousel defined by the peaks of the individual sparks. The resultant
average profile will naturally have two components (similar to those in the
bottom panels of Figure~\ref{fig_p3folds}). As apparent
from Figure~\ref{fig_rfm2}, simultaneous observations of subpulse modulation
in the corresponding pulse-components at two frequencies would exhibit a
phase-offset. Moreover, for a given circulation direction of the carousel,
the observed phase-offset in subpulse modulations at two frequencies in the
leading and trailing components will be equal in magnitude but opposite in sign.
Without loss of generality and for ease of description, unless stated otherwise,
discussion in the rest of the paper would assume a viewing geometry similar to
that shown in Figure~\ref{fig_rfm2} resulting in two pulse components.
\subsection{Formulation}
The subpulse modulation phase, $\Theta$, as a function of
pulse longitude consists of three terms: (1) a scaled version of the
magnetic azimuth, where the scaling factor is the number of
uniformly distributed sparks, \nsp, in the carousel,
(2) an additional, pulse longitude dependent phase term which accounts
for the carousel rotation over the course of the pulse, and
(3) a reference modulation phase at a reference pulse longitude.
The first term also includes the sign of the modulation phase gradient
which depends on the relative direction of the pulsar rotation and the
carousel circulation and is given by sign of $\beta$.
For a known viewing geometry, the magnetic azimuth corresponding
to a pulse phase $\phi$ is given by:
\begin{equation}
\chi (\phi) = \arctan\left(\frac{\sin\zeta\,\sin (\phi-\phizero)}{\sin\alpha\,\cos\zeta - \cos\alpha\,\sin\zeta\,\cos (\phi-\phizero)}\right)
\label{eq_chi1}
\end{equation}
where, $\zeta=\alpha+\beta$, and $\phizero$ is the pulse longitude that
corresponds to the fiducial plane containing the magnetic and rotation axes.
Due to the aberration effects, the conal emission shifts forward by
$r_\nu/\rlc$, where $r_\nu$ is the emission altitude and \rlc\ is the light
cylinder radius \citep{GG03,DRH04}. On the other hand, the core-emission
shifts backward by a similar amount due to retardation effects. In rest of
the paper, \phizero\ has been set to zero assuming that its location is
correctly identified after correcting for the aberration/retardation effects.
Following \citet{ES02}, the subpulse modulation phase is then given by:
\begin{equation}
\Theta(\phi) = -N_{\rm sp}\,\sgn\beta\,\chi(\phi) \,+\, \left(n + \frac{P}{P_3^{\rm obs}}\right)\phi \,+\, \Theta_0
\label{eq_theta}
\end{equation}
where, $P_3^{\rm obs}$ and $n$ are the observed primary modulation period and the
aliasing order, respectively, and $\Theta_0$ is the modulation phase at \phizero.
The actual modulation period, $P_3$, can be
expressed as ${P}/{P_3} = n + {P}/{P_3^{\rm obs}}$, and
$-0.5<{P}/{P_3^{\rm obs}}<0.5$. The circulation
period of the carousel is then $P_4 = N_{\rm sp} \times P_3$.
\par
Assuming the magnetic field to be of dipolar configuration, the magnetic
azimuth for any given field line is a constant. In other words, the
magnetic azimuth
for a given pulse-longitude is independent of the emission altitude.
\textit{Hence, the modulation phase-offset at the peak of a given
pulse-component resulting from simultaneous observations at a pair of
frequencies is entirely decided by the difference in the corresponding
pulse-longitudes and the corresponding change in the magnetic azimuth.}
For aberration-corrected pulse-longitudes of the peaks of a
certain component at two frequencies, $\phia, \phib$, the expected
subpulse modulation phase-offset is:
\begin{equation}
\nonumber
\Delta\Theta = -N_{\rm sp}\,\sgn\beta\,\Delta\chi \,+\, \left(n + \frac{P}{P_3^{\rm obs}}\right) \Delta\phi
\end{equation}
where $\Delta\chi = \chi(\phia)-\chi(\phib)$, and $\Delta\phi=\phia-\phib$.
It is worth mentioning here that \dTheta\ is essentially the
difference between subpulse modulation phases at two different pulse-longitudes,
but observed almost simultaneously at two different frequencies.
The above expression correctly describes the difference in modulation phase
for given pulse longitudes at a given emission height. However, to account
for the emission at different altitudes and to compensate for the carousel
rotation during the corresponding light travel time, ideally an additional
term, $\delta\Theta$, needs to be introduced:
\begin{equation}
\Delta\Theta = -N_{\rm sp}\,\sgn\beta\,\Delta\chi \,+\, \left(n + \frac{P}{P_3^{\rm obs}}\right) \Delta\phi
               + \delta \Theta
\label{eq_dtheta}
\end{equation}
Denoting the emission altitudes at the two frequencies to be $r_{\nu_1}$ and $r_{\nu_2}$,
$\delta\Theta$ can be expressed as:
\begin{equation}
\nonumber
\delta\Theta = \left(n + \frac{P}{P_3^{\rm obs}}\right) \frac{2\pi\left(r_{\nu_1} - r_{\nu_2} \right)}{c\,P}
\end{equation}
For difference of a few hundred kilometers in the emission altitudes, the
light travel time would be only of the order of a millisecond. For a typical
$P_3$ of the order of seconds, $\delta\Theta$ would be negligible in majority
of the cases.
\begin{figure*}
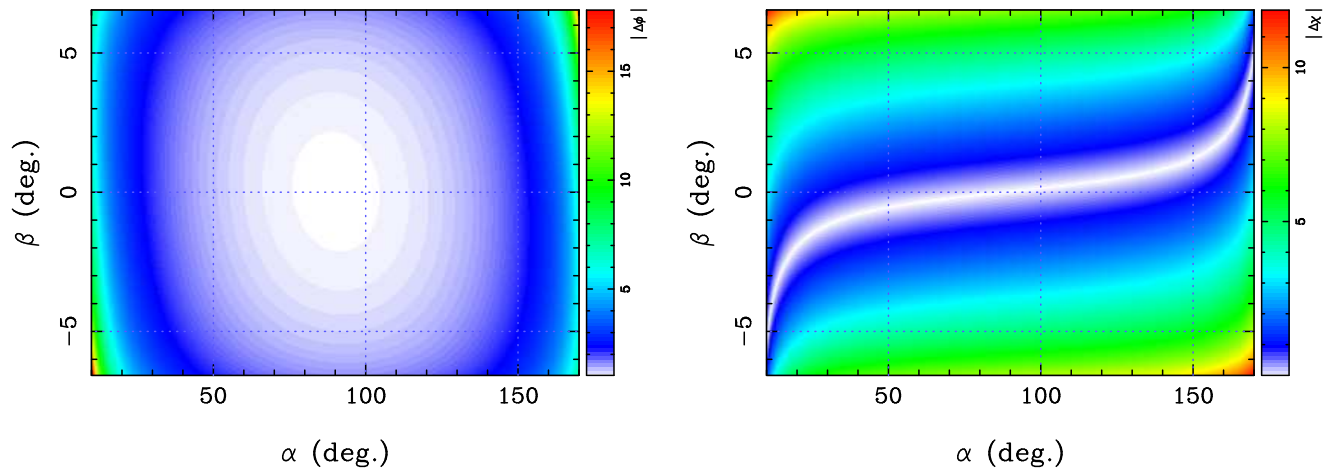

\centering
\subfigure{\includegraphics[width=0.34\textwidth,angle=-90]{delta_absphi_nsp1_bm.ps}
\label{fig_dphi}}
\hspace*{2mm}
\subfigure{\includegraphics[width=0.34\textwidth,angle=-90]{delta_abschi_nsp1_bm.ps}
\label{fig_dchi}}
 \caption{{\sl Left:} Expected change in longitude position (in deg.) of a
pulse-component observed from emission heights 710 and 560\,km (see text for other
details), as a function of $\alpha$ and $\beta$. {\sl Right:} The corresponding
\emph{absolute} variations in the magnetic azimuth (i.e., $\lvert\Delta\chi\rvert$;
in deg.), as a function of the viewing geometry parameters.}
\label{fig_dchi0}
\end{figure*}
\par
The second term in equation~\ref{eq_dtheta} primarily depends on
$\Delta\phi$ and the aliasing order. To estimate the typical expected values
of $\Delta\phi$, we use the geometrical expression for the opening angle of
the emission cone \citep[][]{handbook04}:
$\rho=1.24\mdeg\,(r_{\nu}/10\,km)^{0.5}\,(P/s)^{-0.5}$. For $P$=1.28\,s and
emission heights of 710 and 560\,km corresponding to the assumed frequencies
of 240 and 610\,MHz, the opening angles turn out to be 9.2\mdeg\ and 8.1\mdeg,
respectively. The pulse-longitudes where the sightline cuts with the cone
boundaries on the left side of the fiducial plane at the two emission heights
are assumed to be the peak positions
of the corresponding pulse components, i.e., $\phia$ and $\phib$. For
$-$6.5\mdeg$<\beta<$+6.5\mdeg, 10\mdeg$<\alpha<$170\mdeg, the left panel in
Figure~\ref{fig_dchi0} shows the variations of $\Delta\phi$. For $\alpha$
between about 20\mdeg\ and 160\mdeg, $\Delta\phi$ is only a few degrees. For
such viewing geometries and the assumed emission heights, only very high
aliasing orders ($\lvert n\rvert$$\gtrsim$$5$) would result in the second term to be at
a observationally measurable level. For smaller inclination angles, i.e.,
nearly aligned rotators, $\Delta\phi$ could be 10$-$15\mdeg\ or even more.
For such cases, the contribution of the second term could be significant
even for first aliasing order.
\par
For a reasonably large number of sparks, the first term in
equation~\ref{eq_dtheta} is expected to have dominant contribution in
majority of the cases. For $\Delta\phi$ values in the left panel of
Figure~\ref{fig_dchi0}, the right panel shows the corresponding absolute
changes in the magnetic azimuth (i.e., $\lvert\Delta\chi\rvert$). The
white colored region in this panel ($\Delta\chi$=0) suggests that the points
where the sightline cuts the cone boundaries at the two emission heights
lie on a magnetic radial line. For such viewing geometries, the contribution
of the first term will naturally be zero irrespective of the number of sparks.
For the nearby viewing geometries where $\Delta\chi$ is only a few degrees, a
large number of sparks would help. For all other regions, $\Delta\Theta$
should be easily noticeable even for a few sparks in the carousel.
\par
Figure~\ref{fig_dchi0} provides an estimate of the expected
$\Delta\phi$ and $\Delta\chi$ with the underlying assumptions, however,
the actual quantities would differ depending on the observing frequencies
and the actual emission cone size. For example, $\Delta\chi$ would vary
more steeply with $\beta$ for narrower cones.
\par
Note that the observed $\Delta\Theta$ can also be exploited to constrain the
viewing geometry, especially if $N_{\rm sp}$ is known independently. Particularly
for small inclination angles where $\Delta\chi$, and hence $\Delta\Theta$, is
a steep function of $\alpha$ as well as $\beta$ (Figure~\ref{fig_dchi0}), this
approach can be very useful in constraining the viewing geometry.
\begin{figure*}
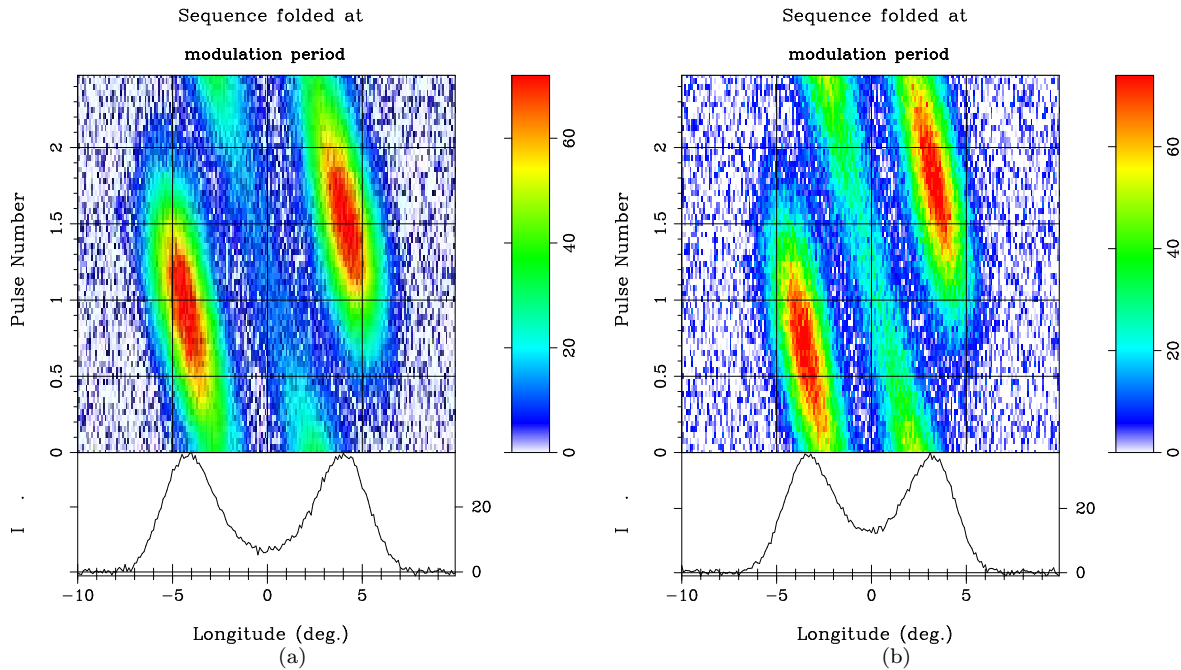

\centering
\subfigure[]{\includegraphics[width=0.47\textwidth,angle=-90]{sm240_modf.ps}
\label{fig_240}}
\hspace*{2mm}
\subfigure[]{\includegraphics[width=0.47\textwidth,angle=-90]{sm610_modf.ps}
\label{fig_610}}
\caption{\ppp-folded modulation patterns for two pulse sequences simulated at
240\,MHz (left) and 610\,MHz (right). The change in separation of the two components
is due to radius to frequency mapping. The vertical change in positions of the
modulation bands under the corresponding pulse components is due to sampling of
the carousel at slightly different rotation phases.}
\label{fig_p3folds}
\end{figure*}
\subsection{Demonstration using simulations}
To demonstrate how the modulation phase-offset would reflect in simultaneous
multi-frequency observations, a carousel pattern similar to that
shown in Figure~\ref{fig_rfm2} but with \emph{6~sparks} was simulated.
The sub-beams were assumed to originate from the sparks uniformly
populated on a 90\,m radius circular ring around the magnetic axis on
the neutron star surface. The intensity distributions of the sub-beams
were modelled as two-dimensional symmetric gaussian functions with half-power
widths nearly 25\% of the carousel radius.
To generate single pulse sequence at a desired frequency, the modulation
sampled by the sightline at the corresponding emission altitude was mapped
to the pulse-longitudes. The pulse sequences at 240 and
610\,MHz were generated by assuming a carousel circulation period of 14.84\,P
(and hence, a subpulse modulation period P$_3$ of 2.47\,P).
Furthermore, normally distributed noise was added to the
pulse-sequences such that the average signal-to-noise ratio (S/N) of
resultant individual pulses was nearly 10.
\par
The different sampling of the carousel would result in the corresponding
``drift-bands'' at two frequencies to appear offset from each other. This
offset in the subpulse modulation phase can be best viewed after folding
the sequences modulo \ppp\ (and hence increasing the significance of the
drift-bands). The \ppp\ modulo folded pulse sequences, hereafter called as
\ppp-folds, are shown in Figure~\ref{fig_p3folds}. The drift-bands or
more generally, the modulation bands, i.e., the modulation patterns under
the individual pulse-components, are clearly offset at two frequencies.
The larger horizontal separation between the drift-bands at the lower
frequency is due to RFM. The vertical offset between the corresponding
modulation bands at the two frequencies is a consequence of slightly
different sampling of the carousel at the polar cap. As expected, the
vertical offsets are in opposite directions for the two components.
By cross-correlating the modulation patterns at the peak positions
of the leading component at the two frequencies (at about $-$4.1\mdeg\
and $-$3.3\mdeg\ of pulse-longitude), the offset
is estimated to be about 0.11\,\ppp\ (or 0.26\,$P$). Note that this offset
is primarily due to the first term in equation~\ref{eq_dtheta}, with the
second and third terms contributing only about 0.8\% and 0.1\%, respectively.
An example of this effect on real data from PSR~B0943+10 can be seen in
figure~14 of \citet[][]{Bilous18}.
\subsection{Implications for determining carousel related physical parameters}
\label{sect_params}
The \emph{pulse-to-pulse} intensity fluctuation sequence corresponding to a
particular component at
a given frequency could be considered as a delayed or advanced version of that
at a different frequency since the corresponding sightlines sample the
underlying carousel at its different rotation phases. An expression equivalent
to equation~\ref{eq_dtheta} can be written for the absolute delay between
fluctuations under a component at two frequencies. Considering only the first
term in equation~\ref{eq_dtheta}, a simplified expression for this delay,
$\Delta t$, is:
\begin{equation}
\Delta t = \frac{\Delta\chi}{\omega_c}
\label{eq_dtobs}
\end{equation}
where, $\omega_c$ is the angular velocity of the carousel\footnote{For
situations where the contributions from the second and third terms in
equation~\ref{eq_dtheta} are significant, the corresponding terms should
be incorporated in equation~\ref{eq_dtobs}.}. Measurement of
$\Delta t$ would provide direct estimates of the angular velocity of the
carousel (and hence \pppp) and its circulation direction. Using the
viewing geometry parameters and independently measured emission altitude,
$\omega_c$ can be easily translated to the drift velocity of the sparks.
Note that this method can provide a direct estimate of the the \emph{average}
drift velocity even if the underlying carousel is not stable enough to give
rise to observable subpulse modulation. The drift velocity directly relates
to the electric potential in the polar gap in the carousel model \citep{RS75},
and provides important feedback to the pulse emission models.
\subsubsection{Resolving the aliasing of \ppp}
The shortest \ppp\ that can be
measured without getting aliased is about twice the pulsar rotation period.
An estimate of \pppp\ in combination with the $P_3^{\rm obs}$ can potentially
constrain the aliasing. As shown by \citet{DR01}, solving for integer
pairs of n and $N_{\rm sp}$ in $P_4 = N_{\rm sp}/\left(n + {P}/{P_3^{\rm obs}} \right)$
along with observed polarization constraints could provide the aliasing
order. However, measuring \pppp\ reliably and using it to resolve the
\ppp-aliasing heavily depends on stability timescale of the carousel.
\par
The estimate of $\Delta t$ using simultaneous dual- or
multi-frequency observations enables an independent and
efficient way of resolving the \ppp-aliasing. For a reasonably coherent subpulse
modulation, the phase-offset in modulation at two frequencies under a specific
component can be measured using, e.g., the \ppp-folds (similar to those presented
in Figure~\ref{fig_p3folds}). This \emph{measured} phase-offset is related to
the measured delay via:
\begin{equation}
\dTheta_{\rm obs} = 2\pi\left(n + \frac{P}{P_3^{\rm obs}}\right) \frac{\dtobs}{P}
\label{eq_alias}
\end{equation}
where the subscript or superscript ``obs'' denotes that these are observational
estimates.
The estimates of \dTheta$_{\rm obs}$, $P_3^{\rm obs}$ and \dtobs\
can be readily used to determine the permissible aliasing order. It is worth
emphasizing here that equation~\ref{eq_alias} involves \emph{only} observed
quantities and does not depend on viewing geometry parameters. Moreover,
this method is also independent of the shape or symmetry axis of the
carousel, and can reliably be employed even for a partial carousel. A
relevant demonstration can be found in Section~\ref{sect_b1237}.
\section{Implications for carousel related observable phenomena}
There are a variety of observed phenomena that are either considered
to be generally associated with the carousel of sparks at the polar cap,
or they are examined within the scope of the carousel model. The implications
of the sightline sampling different parts of the carousel at different
frequencies are discussed below in the context of some of these phenomena.
\subsection{Multi-altitude emission at same frequency}\label{mae}
Majority of pulsars exhibit average profiles with 5 or lesser number
of components. These components are explained by the presence of a single
or two nested hollow emission cones centered at the magnetic axis and a
`core'-beam near the axis \citep{Rankin83a}. However, the co-existence of
two emission cones would indicate that of the corresponding presumably two
concentric carousels in the polar cap region. \citet{Rankin93a} suggested
an interesting possibility that the emission apparent from the inner cone
happens at a lower height along the same group of peripheral field lines
that produce the outer cone emission components. In this ``single frequency
multi-altitude emission'' scenario, a single emission cone corresponding to
a single carousel of sparks could produce multiple components in the
observed profile.
\par
\citet{GG01} measured the emission heights separately for different
components of the bright pulsars PSR~B0329+54 and proposed that the
different components originate at different heights in the magnetosphere
but along relatively nearby field lines. For PSR~B1237+25,
\citet[][hereafter MD14]{MD14}
showed that the underlying emission patterns corresponding to the presumably
inner and outer cones are significantly correlated with each other. They
proposed that the same radio frequency emission in the two cones originates
from a common seed pattern of sparks at two different altitudes. These two
case studies have strengthened the hypothesis of multi-altitude emission
in pulsars.
\par
The multi-altitude emission scenario implies that components from presumed different
emission cones sample the carousel at its multiple rotation phases even in a
single frequency observation. Hence, as discussed in previous sections, the
components from different cones should exhibit a modulation phase-offset in
the multi-altitude scenario. Moreover, the phase-offset is expected to have a
sign reversal when examined for a pair of components originated on the other
side of the fiducial plane, and provides an important test for the single
frequency multi-altitude proposition. Furthermore, confirming the modulation
phase-offsets and their opposite signs for the pairs on the two sides of the
fiducial plane does not require any knowledge about the viewing geometry.
For a known emission geometry, the measured phase-offsets can be used to
measure carousel parameters as explained in
Section~\ref{sect_params}. A detailed case study to test the multi-altitude
emission scenario is presented in Section~\ref{sect_b1237}.
\subsection{Magnetic field twisting}\label{mft}
A progressive twist in the magnetic field geometry as we go in the
relatively weaker field regions further away from the star has been
speculated \citep[][\citetalias{MD14}]{RSD03}. One of the ways to probe any twist in the
field lines is to map the underlying emission patterns (in the form of a
carousel of emission sub-beams) at different frequencies (i.e. at different
altitudes) and examining any rotation of the pattern \citep[][\citetalias{MD14}]{Maan13}.
While this method makes use of the underlying modulation and advantages
of averaging, it suffers from most likely uneven distribution of modulation
power in different components and could be misleading. Most of the pulsars
have asymmetric profiles or with significantly different intensity under
different pulse components. The underlying modulation power also varies
significantly under different pulse components. The corresponding mapping
of the emission patterns naturally gets dominant contributions from the
components with stronger modulation powers. Along with the frequency or emission
height dependent variation in modulation phase discussed above, the asymmetric
or uneven distribution of modulation power could result in an apparent rotation
of the mapped emission patterns even in the absence of any twist in the field
geometry.
\par
Examining the effects of field twisting on \dTheta\ would provide a cleaner
probe. In the presence of a twist in field lines, equation~\ref{eq_dtheta}
would get contribution from an additional term, $\delta\Theta_{\rm twist}$,
which would be a measure of the twist between the two emission altitudes under
consideration. Importantly, $\delta\Theta_{\rm twist}$ contribution will be
in the same direction (i.e., without a sign reversal) on both sides of the
fiducial plane. This aspect can be used to measure any twist in the magnetic
field lines between two emission altitudes. If we denote the phase-offsets
measured on the two sides of the fiducial plane as \dTheta$_1$ and \dTheta$_2$,
then
\begin{equation}
\delta\Theta_{\rm twist} = \frac{\dTheta_1+\dTheta_2}{2}
\end{equation}
and the corresponding change in the magnetic azimuth of a
field-line between the two emission heights can be expressed as
$\delta\chi_{\rm twist}$
\begin{equation}
\delta\chi_{\rm twist} \approx \frac{\delta\Theta_{\rm twist}}{N_{sp}}
\end{equation}
\subsection{Pseudo-nulls observed at multiple frequencies}
Pseudo-nulls represent chance positioning of our sightline across the minima
(i.e., in between the sub-beams) in the carousel \citep{HR07,HR09}, as against
the physical cessation of the emission in the actual nulls. Due to the above
discussed frequency dependent phase-offset, pseudo-nulls can be expected to
be non-simultaneous for carefully chosen observing frequencies. Since the
actual nulls are generally broadband \citep{Gajjar14}, the above aspect
provides a critical test for the presence of pseudo-nulls.
\subsection{Wide-band observations are not necessarily more sensitive
for single pulse modulation studies}
A pulse-sequence from a significantly wide-band observation would merge the
resultant modulation from a finite \emph{angular width} of the carousel.
If multiple frequencies were to sample the same part of the carousel, RFM
would have implied observing the same subpulse at different pulse-longitudes
as the position of a pulse-component changes with frequency, which in turn,
would imply the drift-bands to be \emph{wider} in pulse-longitude for wide-band
observations. However, as the corresponding subpulse modulation phase also
changes with frequency, wide-band observations imply the observed drift-bands
to get vertically \emph{elongated}. In case the corresponding change in
pulse-width across the observed frequency range is smaller or comparable to
the subpulse-width, the merging of subpulses at different rotation phase of
the carousel implies that the subpulse modulation would get smeared, and might
even be washed out for appropriate combinations of viewing geometry, \ppp, and
observing bandwidth.
\par
In addition to the frequency dependent phase-offset itself, its manifestation
in some of the above discussed phenomena would provide critical tests of the
carousel model.
\begin{figure*}
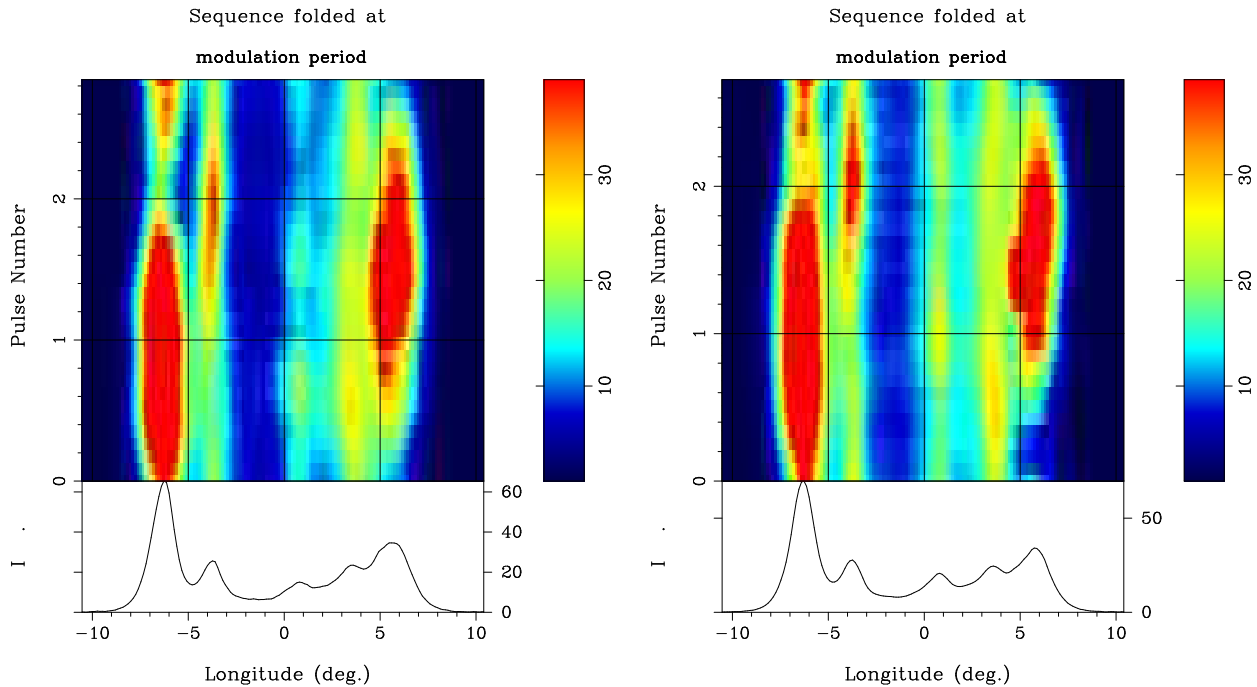

\centering
\subfigure{\includegraphics[width=0.5\textwidth,angle=-90]{modf_97_352fpol.ps}
\label{b1237_p3folds1}}
\hspace*{2mm}
\subfigure{\includegraphics[width=0.5\textwidth,angle=-90]{modf_417_672fpol.ps}
\label{b1237_p3folds2}}
\caption{\textsl{Left:} The subsequence~\s1\ (pulse-number range 97--352)
folded modulo \pppobs=2.84\,P. \textsl{Right:} The \ppp-folds for subsequence~\s2
(pulse-number range 417--672) corresponding to it's \pppobs\ of 2.72\,P.
The colors have been saturated a little to bring out the modulations clearly.
Presence of modulation bands for all the four conal components, including
component IV, are clearly visible for both the subsequences. The \ppp-folds
were computed with reference to arbitrary starting pulses, however, the
reference pulses were tried to be adjusted so as to get the patterns for
the two subsequences nearly aligned for ease of comparison.}
\label{b1237_p3folds}
\end{figure*}
\section{Multi-altitude emission and resolving \ppp--aliasing: a case study
using PSR~B1237+25}\label{sect_b1237}
The bright pulsar PSR~B1237+25 exhibits a multi-component profile resulting
from a special viewing geometry and a rich variety of subpulse modulation
in many of its components. The impact angle estimate 
\citep[$|\beta|\lesssim 0.25\mdeg$;][]{SR05} indicates that our sightline
traverses almost through the magnetic axis of this pulsar. The sightline
presumably cuts through two hollow emission cones and a core beam giving
rise to five components in the average profile (for example, see lower panels
in Figure~\ref{b1237_p3folds}).
For ease of reference in the discussion below, let us denote the five
distinct components, starting from the leading most one as I, II, III, IV
and V. Component I and V are the \emph{outer conal} components, II and IV
are the \emph{inner conal} components, and III is the \emph{core component}.
The emission
height estimates for the inner and outer cones using aberration-retardation
considerations \citep[$\sim$280 and 340~km;][]{SR05} and those after taking
care of even the sweep-back effects of the field lines
\citepalias[$\sim$640 and 715~km;][]{MD14}, suggest the inner cone to be emitted
at altitudes lower than the outer cone.
\subsection{Single frequency multi-altitude emission}
Various characteristics of B1237+25 were indeed some of the compelling
reasons that \citet{Rankin93a} proposed the multi-altitude emission hypothesis.
Based on the correlation between the emission patterns corresponding to
the two cones and the emission
height estimates, \citetalias{MD14} provided a strong support to the
\emph{multi-altitude} scenario for this pulsar.
Given that the outer conal components
and at least one of the inner conal components exhibit rich variety of
subpulse modulation \citep[e.g., see][\citetalias{MD14}]{SR05},
B1237+25 offers an interesting but challenging case to test the
multi-altitude emission proposition by examining the modulation phase
offsets in components corresponding to the two cones (cf. Section~\ref{mae}).
\par
To test the multi-altitude emission scenario as well as to explore the
scope of resolving the aliasing issue using methods proposed in
Section~\ref{sect_params}, a detailed analysis of a 5209 rotation periods
long pulse sequence from this star observed using the Arecibo telescope at
327\,MHz is presented here. The sequence has a time resolution equivalent
to 0.133\mdeg\ in pulse-longitude. More details of the observations as well
as basic data reduction (e.g., dedispersion, Faraday rotation correction,
etc.) can be found in \citet{SRM13} and \citetalias{MD14} (sequence ``A'' therein).
\par
As expected from a central sightline traverse, the pulse-to-pulse intensity
fluctuations
in B1237+25 are primarily of the form of \emph{amplitude modulation} with
\pppobs\ around 2.7--2.8\,P. Moreover, the subpulse modulation does not
seem to be stable over long durations, perhaps due to a temporally unstable
carousel. This results in a broad frequency feature corresponding to \pppobs\
(hereafter the \ppp-feature) in the longitude-resolved fluctuation spectrum
\citep[LRFS;][]{Backer70c}. To determine the modulation phase-offsets and test
the multi-altitude emission proposition, all the four conal components need to
exhibit the same modulation. However, noting from the previous studies,
component~IV hardly shows any indication
of modulation. Presence of any faint modulation in this component is perhaps
best sought in small sections of the sequence that could potentially exhibit
a coherent modulation.
To find instances of stable and coherent modulation at shorter timescales,
a number of LRFS computed using 256-pulse wide blocks, with the starting
of successive blocks separated by 16 pulses, were examined. Two blocks, with
the pulse-number ranges 97--352 and 417--672 (hereafter subsequence \s1 and \s2,
respectively), were identified to have particularly strong and reasonably
coherent \ppp-features. The corresponding \pppobs\ in these two subsequences were
measured to be 2.84\,P and 2.72\,P, respectively. These subsequences folded
modulo their respective \pppobs\ are shown in Figure~\ref{b1237_p3folds}.
\par
\begin{deluxetable}{@{\extracolsep{4pt}}ccccc}
\tabletypesize{\footnotesize}
\tablecolumns{5}
\tablewidth{0pt}
\tablecaption{Cross-correlation and phase-offsets between \ppp-folded
modulation patterns.}
\tablehead{
           \colhead{Subseq.}                                           &
           \multicolumn{2}{c}{Components I and II}                      &
           \multicolumn{2}{c}{Components IV and V}                      \\
           \cline{2-3} \cline{4-5}
           \colhead{}                                                   &
           \colhead{$\rho_{\rm max}$ (\%)}                              &
           \colhead{$\Delta \Theta_{\rm obs}$(\mdeg)}                    &
           \colhead{$\rho_{\rm max}$ (\%)}                              &
           \colhead{$\Delta \Theta_{\rm obs}$(\mdeg)}                    
          }

\startdata
\s1 & $91.0\pm6.7$  & $+137\pm32$ & $76.5\pm16.4$   & $-104\pm36$  \\
\s2 & $95.6\pm3.4$  & $+137\pm22$ & $71.3\pm19.0$   & $-115\pm47$  \\
\enddata
\tablecomments{$\Delta \Theta_{\rm obs}$ is the phase-offset by which the \ppp-folded
modulation-patterns under the inner conal components II and IV \emph{lag} behind
those under the outer conal components I and V, respectively, and it corresponds
to the maximum normalized cross-correlation coefficient, $\rho_{\rm max}$.}
\label{tab1}
\end{deluxetable}
\par
There are quite a few things to notice in the P3-folds presented in
Figure~\ref{b1237_p3folds}. First of all, the coherency of the \ppp-feature
in LRFS reflects in the form of clearly identifiable modulation bands for
components I, II and V. Despite of having slightly different \pppobs\ values,
the overall modulation bands are similar for both the subsequences.
The component IV generally does not show a distinct \ppp-feature in LRFS,
but Figure~\ref{b1237_p3folds} clearly shows that this component also exhibits
modulation at the same \pppobs. The modulation band for component IV is less
prominent than those for the other three components, perhaps because
of a significant overlap with component V and since the modulations in
these two components are clearly offset in phase. The modulation bands in 
components I and II are also offset, and interestingly, the offset appears
to be similar as that between components IV and V but in opposite direction.
Finally, the core component does not show much indication of a well defined
modulation band.
\par
To measure the modulation phase-offsets, the modulation patterns corresponding
to the peaks of the inner and outer conal components (after averaging over a
few bins around
the peak) were extracted from the \ppp-folds and cross-correlated with each other.
Measurements of the maximum normalized cross-correlation coefficients as well as
the corresponding phase-offsets ($\Delta\Theta_{obs}$) are given in Table~\ref{tab1}.
Note that the
measured correlation between the modulation patterns is highly significant. 
The larger uncertainties for the pair of components IV and V are perhaps due
to the significant overlap between the two components. Modulation in component II
lags behind that in I, while the modulation in component IV leads that in V
by similar amounts within the measurement uncertainties. It was also confirmed
that the modulation patterns, and hence the phase-offset measurements, are
not affected by any profile or polarization mode changes (see Appendix).
\par
The above measurements clearly show that the phase-offsets in modulations under
the inner and outer conal component pairs on the two sides of the fiducial plane
are in agreement with the expectations from a multi-altitude emission scenario.
Noting again that (i) the inner cone is emitted at a lower altitude than the
outer cone, (ii) the modulations in the inner and outer conal components are
phase-locked and exhibit same \ppp, and (iii) the phase-offsets in these
corresponding modulations are equal and opposite on the two sides of the
fiducial plane, it seems almost inevitable to conclude that multi-altitude
emission along nearby bunches of field lines seeded by a single carousel of
sparks in the polar cap gives rise to the observed average and modulation
properties of B1237+25.
\par
The ``modified carousel model'' \citep{GS00} proposes maximal packing of
the sparks on the polar cap and can explain the subpulse modulation in
multiple cones originating from multiple carousels. Due to a characteristic
spark dimension that is similar to separation between consecutive sparks, this
model predicts the number of sparks in the inner carousel to be less
than those in the outer carousel. However, at least for two double-cone pulsars
B0818$-$41 \citep{BGG09} and B1237+25 \citepalias{MD14}, the number of subbeams
in the two apparent carousel rings have been found to be same.
Hence, the multi-altitude emission is perhaps the only viable explanation
for the observed properties in double-cone pulsars. 
Given the similar observed properties, including the
phase-locked modulation \citepalias[also see detailed discussion in][]{MD14},
for many other multiple component pulsars, the above modulation phase analysis
could be used to confirm the multi-altitude emission to be, perhaps, a common
phenomena in pulsars.
\par
\citetalias{MD14} also observed magnetic azimuth offsets between
the carousel patterns mapped for the inner and outer conal component pairs.
They associated this offset to a potential twist in the emission columns
(and hence the magnetic field geometry) between the two different emission
altitudes. However, as apparent from their figure 2, the subpulse modulation
power under component II heavily dominates that under component IV (it is
also apparent from Figures~\ref{b1237_p3folds} and \ref{b1237_p3foldsppm}
in this work). Hence, as explained in Section~\ref{mft}, the analysis in
\citetalias{MD14} to determine any twist could have been affected by the
highly uneven distribution of modulation powers under the two inner
conal components. If that is indeed the case, the magnetic azimuth shift
measured in \citetalias{MD14} multiplied by the corresponding number of
sparks would be a quantity directly comparable to $\Delta\Theta_{obs}$ in
this work. Using their tables 1 and 2, this quantity is 180\mdeg$\pm$25\mdeg,
156\mdeg$\pm$42\mdeg, 117\mdeg$\pm$45\mdeg and 80\mdeg$\pm$36\mdeg,
corresponding to the four subsequences presented therein, and is largely
consistent with $\Delta\Theta_{obs}$ within the measurement uncertainties
(Table~\ref{tab1}). Hence,
the magnetic azimuth offsets measured in \citetalias{MD14} could more likely be
explained as the modulation-phase offsets between components I and II\footnote{In
carousel mapping, the modulations in the outer conal components I and V would get
added in phase, and the modulation power in II dominates that in IV.}
--- a natural consequence of multi-altitude emission and the
carousel model rather than a twist in the emission columns.
However, the measurements in Table~\ref{tab1} might give an impression of small
difference in the phase-offsets on the two sides of the fiducial plane, which
would hint towards a small twist ($\delta\chi_{twist}=[16\mdeg\pm24\mdeg]/N_{sp}$;
much smaller than that in \citetalias{MD14} for number of sparks more than a few)
in the field lines (Section~\ref{sect_params}).
But within the current measurement uncertainties, this analysis does not
present an evidence supporting the field twist. A more detailed analysis
using lower frequency observations, where the outer and inner conal components,
particularly components IV and V, would perhaps be better separated, is under
progress.
\begin{deluxetable}{@{\extracolsep{4pt}}ccccc}
\tabletypesize{\footnotesize}
\tablecolumns{5}
\tablewidth{0pt}
\tablecaption{\emph{Absolute} delays between pulse-to-pulse intensity
fluctuations under different components.}
\tablehead{
           \colhead{Subseq.}                                           &
           \multicolumn{2}{c}{Components I and II}                      &
           \multicolumn{2}{c}{Components IV and V$^{\dagger}$}         \\
           \cline{2-3} \cline{4-5}
           \colhead{}                                                   &
           \colhead{$\Delta t_{\rm obs}$ (P)}                           &
           \colhead{$\Delta \Theta_{\rm imp}$(\mdeg)}             &
           \colhead{$\Delta t_{\rm obs}$ (P)}                           &
           \colhead{$\Delta \Theta_{\rm imp}$(\mdeg)}             
          }

\startdata
\s1              & $+1.09\pm0.14$  & $+137\pm18$ & $-0.55\pm0.40$   & $-68\pm50$  \\
\s2$^{\ddagger}$ & $+1.30\pm0.35$  & $+173\pm47$ & $-0.52\pm0.50$   & $-68\pm65$  \\
\enddata
\tablecomments{The delay ($\Delta t_{\rm obs}$) measurements indicate how much
the pulse-to-pulse intensity fluctuations under the inner conal components II
and IV \emph{lag} behind those under the outer conal components I and V,
respectively. $\Delta \Theta_{\rm imp}$ is the phase-offset \emph{implied} by
the measured delay, i.e., $\Delta \Theta_{\rm imp}={360\mdeg\times\Delta t_{\rm obs}/\ppp}$,
assuming the aliasing order to be 0.}
\tablenotetext{$\dagger$}{The significant overlap between the components IV and V
makes it harder to measure the underlying phase-gradient. Although a definitive
phase-gradient and corresponding delay has been measured, it is possible that
the overlap of the two modulations has contaminated the measurements, see text
for more discussion.}
\tablenotetext{$\ddagger$}{These measurements have utilized only first 100 pulses
in subsequence~\s2 to avoid the contamination due to abnormal mode instance.}
\label{tab2}
\end{deluxetable}
\subsection{\ppp-aliasing and carousel parameters}
With the above supporting evidence for multi-altitude emission, the
\emph{pulse-to-pulse} intensity fluctuations under the outer conal components
can be viewed as delayed versions of those under the corresponding inner conal
components. The delay, if measurable, can be used to resolve the \ppp-aliasing
as discussed in Section~\ref{sect_params}.
In principle, the single pulse intensity fluctuations corresponding
to different components can be cross-correlated to estimate any underlying delay.
However, such an approach would be severely limited by our sampling of the
fluctuations only once every rotation period. Instead, we can compute the
cross-spectrum of two fluctuation sequences of interest and look for any
significant gradient in the cross-spectrum phases. This technique can be
used to measure any fractional delays with better precision, and is often
used in antenna and digital signal processing. Using this method, the
measurements of fractional delays ($\Delta t_{obs}$) between fluctuation
sequences under the above discussed pairs of components are presented in
Table~\ref{tab2}.
\par
As evident from Table~\ref{tab2}, the delay between fluctuations under
components I and II is highly commensurate with the corresponding
phase-offset measurements in Table~\ref{tab1}. The delay measurements
for the component pair IV and V are also consistent with the corresponding
phase offsets within the uncertainties, though there is a hint that the
delays are underestimated.
The modulations in the components IV and V are offset by more than a quarter
of \ppp-cycle, and their significant overlap is prone to contaminate the resultant
phase gradient in the cross-spectrum. Since the underlying fluctuation
spectrum is same, the overlap of two out-of-phase modulations under these
two components would result in underestimation of the actual phase gradient,
and hence that of the corresponding delay\footnote{Note that, in this
situation, the cross-correlation of the corresponding \ppp-fold patterns (as
in Table~\ref{tab1}) would still provide \emph{less} contaminated estimates of
$\Delta\Theta$ as long as the intrinsic modulation powers in the individual
components dominate the corresponding modulation patterns.}. Due to these
reason, below we use the delay estimates only for the components I and II
to determine the aliasing order.
\par
Using the $\Delta\Theta_{obs}$ and $\Delta t_{obs}$ measurements for the
pair of components I and II (Tables~\ref{tab1} and \ref{tab2}) in
equation~\ref{eq_alias} results in the aliasing order estimates of
0.0$\pm$0.09 and $-$0.08$\pm$0.09 for the subsequences $S_1$ and $S_2$,
respectively. Since $n$ has to be an integer, the only permissible
solution is 0. Note that the immediate next aliasing orders
($\lvert n \rvert$=1) are highly inconsistent with the above estimates. Hence,
\emph{within the multi-altitude emission scenario}, it is fairly evident
that \pppobs\ of 2.7--2.8\,P is the actual \ppp.
\par
Now equation~\ref{eq_dtheta} can also be used to estimate the number
of sparks. Since the $\Delta\Theta_{obs}$ estimates for the component pair IV and V
may be affected by their overlap, only the corresponding estimates between
the first two components will be used below.
To compute $\Delta\phi$ and $\Delta\chi$, first the positions of
the components are needed. \citet{SR05} provide (in their table 1) the
positions of the conal components as well as the aberration-retardation
longitude-shifts\footnote{Independently estimated positions by fitting the
average profile using a number of Gaussian functions were found to be
consistent with those from \citet{SR05}.}. After correcting for the
aberration-retardation shifts, the positions of the components I and II
are $-$6.03\mdeg$\pm$0.22\mdeg\ and $-$3.66\mdeg$\pm$0.13\mdeg, and hence
$\Delta\phi$=2.37\mdeg$\pm$0.26\mdeg. Using $n$=0, the second term in
equation~\ref{eq_dtheta} contributes only about 0.6\% of the observed
$\Delta\Theta$ of 137\mdeg\ (Table~\ref{tab1}). The contribution from the
third term is expected to be even smaller. Using the above component positions
in equation~\ref{eq_chi1}, $\Delta\chi$ is estimated to be 1.52\mdeg$\pm$0.25\mdeg\
for $\alpha$=$57.6\mdeg$\ and $\beta$=$-0.3\mdeg$\ \citep{SRM13}.
Equation~\ref{eq_dtheta} then suggests the number of sparks to be some
60--133. If we also include an uncertainty of $\pm$0.1\mdeg\ in $\beta$,
this estimate becomes even less constraining to 40--240. The corresponding
estimate for the carousel circulation period would be in the range
113--680\,P. Using $\Delta t_{obs}$ estimates in equation~\ref{eq_dtobs}
also provides similar estimates.
Better constraints on the number of sparks as well as \pppp\ will require
more precise estimates of $\Delta\Theta_{obs}$ and $\beta$.
\section{Conclusions}
Simultaneous multi-frequency observations have been utilized to
probe and characterize several properties of pulsars. However, observable
geometric effects of radius-to-frequency mapping in the subpulse
modulations manifested by the underlying carousel of sparks have largely
remained unexplored. This work presents the conceptual as well as mathematical
formulation of the expected geometric imprints of the carousel in simultaneous
multi-frequency or wideband observations. The main conclusions of this work
are summarized as follows.
\begin{enumerate}
\item A formulation has been presented for the expected change in modulation
phase under a pulse component observed simultaneously at two or more frequencies.
The phase-offset is illustrated using simulated pulse sequences at two
frequencies, and it's expected variations with the viewing geometry are
demonstrated with a reasonable assumption about the frequency dependence
of the emission cone size.
\item It is shown that the above phase-offset can be used to resolve the
aliasing issue in the measured \ppp\ without relying on knowledge of the
viewing geometry of the pulsar. For a known emission geometry,
the phase-offset can also be used to determine various physical parameters
associated with the carousel, e.g., the average velocity of the sparks and
the carousel circulation period.
\item The phase-offset between subpulse modulations observed simultaneously
at two or more frequencies has important implications in probing several
phenomena (e.g., single frequency multi-altitude emission, field twisting,
orthogonal polarization modes emission sites, etc.) as well as scrutinizing
the carousel model and related phenomena (e.g., the pseudo-nulls).
\item A detailed analysis of a 327\,MHz pulse sequence from B1237+25 has
been presented uncovering the modulation even under the inner conal component
on the trailing side. The phase-offsets between modulations under the inner
and outer conal components are presented as a strong evidence for single-frequency
multi-altitude emission in B1237+25. \textit{Within this multi-altitude scenario,}
it is shown that the measured \ppp\ of 2.7--2.8\,P is definitely not aliased.
\end{enumerate}
The formulation, methods and analysis presented here can be readily applied to
already existing as well as future simultaneous multi-frequency or wideband
observations to scrutinize pulsar emission models and provide further
observational feedback.
\vspace*{2mm}
\acknowledgments
I sincerely thank the anonymous referee for constructive comments which helped in
improving the manuscript significantly.
I am grateful to Joanna Rankin for making available the 327\,MHz pulse sequence
discussed in Section~\ref{sect_b1237} as well as its polarization mode separated
version used in the Appendix. I also sincerely thank Joeri van Leeuwen and
Andrzej Szary for their valuable feedback on the manuscript.
A part of the work presented here made use of a modified version of the single
pulse analysis software \texttt{SPULSES}, which was originally developed by
Prof. Avinash A. Deshpande at the Raman Research Institute, Bangalore, India.
I acknowledge use of the funding from the European Research Council under the
European Union's Seventh Framework Programme (FP/2007-2013)/ERC Grant Agreement
no. 617199.
\appendix
\begin{figure*}
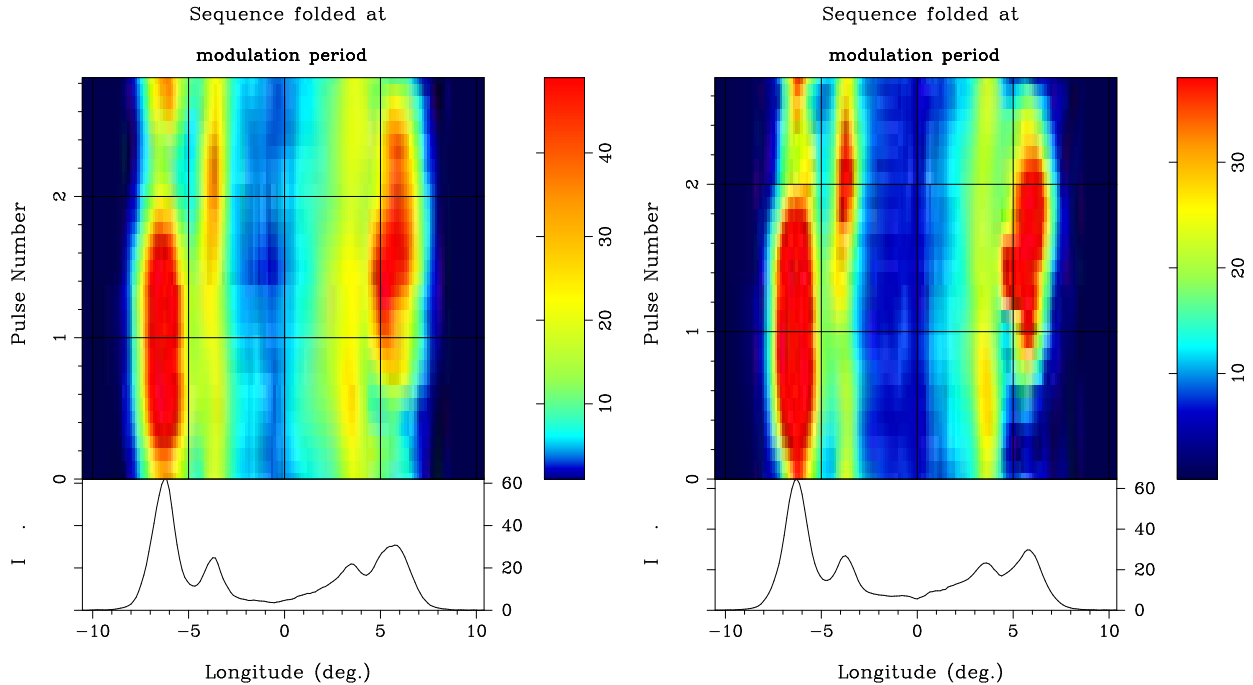

\centering
\subfigure{\includegraphics[width=0.5\textwidth,angle=-90]{modf_97_352rotated.ps}}
\hspace*{2mm}
\subfigure{\includegraphics[width=0.5\textwidth,angle=-90]{modf_417_672.ps}}
\caption{Same as in Figure~\ref{b1237_p3folds}, but for the PPM powers from
the subsequence \s1 (left) and \s2 (right).}
\label{b1237_p3foldsppm}
\end{figure*}
Like several other pulsars, B1237+25 exhibits two modes: normal and abnormal
modes \citep{Backer70a,Backer70b,Backer70c,Backer73}, which are marked by
significantly different profile shapes as well as
the relative orthogonal polarization mode (hereafter OPM) powers. The classical
normal mode was later found to consist of two different sub-modes: a flare-normal
mode defined by bright core activity and a quiet-normal mode with little or
no core activity \citep{SR05}. The subpulse modulation properties could differ
in different modes. However, we note that there is no noticeable instance of
abnormal mode pulses in subsequence~\s1, while subsequence~\s2 shows one
transition to the abnormal mode for about 20 periods. By examining the
\ppp-folds using only the normal mode pulses before and after the abnormal
mode instance, it was confirmed that the \ppp-folds presented in
Figure~\ref{b1237_p3folds} are not affected by the mode-change in
subsequence~\s2. As proposed by \citet{RR03}, the emission elements
giving rise to the modulating
powers in the two OPMs, namely the primary polarization mode (PPM) and the
secondary polarization mode (SPM), could also have geometrical offsets.
Such offsets, if any, could smear the modulation bands in Figure~\ref{b1237_p3folds}.
To examine the possible presence of any such effects, a two-way polarization
separated version of the pulse sequence was used to obtain the \ppp-folds
for the two subsequences using only the PPM power. These \ppp-folds are presented in
Figure~\ref{b1237_p3foldsppm} and clearly show that the
modulation bands in the conal components are identical to those obtained
using the total power in Figure~\ref{b1237_p3folds}. The core-component mostly
consist of SPM power, and hence, it is nearly absent in Figure~\ref{b1237_p3foldsppm}.
If at all, the near absence of core component in Figure~\ref{b1237_p3foldsppm}
makes it further easier to notice
the phase-offset between the modulations under component IV and V.


\begin{thebibliography}{}
\expandafter\ifx\csname natexlab\endcsname\relax\def\natexlab#1{#1}\fi
\providecommand{\url}[1]{\href{#1}{#1}}

\bibitem[{{Asgekar} \& {Deshpande}(2001)}]{AD01}
{Asgekar}, A., \& {Deshpande}, A.~A. 2001, MNRAS, 326, 1249

\bibitem[{{Asgekar} \& {Deshpande}(2005)}]{AD05}
---. 2005, MNRAS, 357, 1105

\bibitem[Backer(1970a)]{Backer70a}
Backer, D.~C.\ 1970a, Nature, 228, 42
%
\bibitem[Backer(1970b)]{Backer70b}
Backer, D.~C.\ 1970b, Nature, 228, 752
%
\bibitem[Backer(1970c)]{Backer70c}
Backer, D.~C.\ 1970c, Nature, 228, 1297
%
\bibitem[Backer(1973)]{Backer73}
Backer, D.~C.\ 1973, ApJ, 182, 245
%
\bibitem[{{Backus} {et~al.}(2010){Backus}, {Mitra}, \& {Rankin}}]{BMR10}
{Backus}, I., {Mitra}, D., \& {Rankin}, J.~M. 2010, MNRAS, 404, 30

\bibitem[{{Bhattacharyya} {et~al.}(2009){Bhattacharyya}, {Gupta}, \&
  {Gil}}]{BGG09}
{Bhattacharyya}, B., {Gupta}, Y., \& {Gil}, J. 2009, MNRAS, 398, 1435

\bibitem[Bilous(2018)]{Bilous18}
Bilous, A.~V.\ 2018, A\&A, 616, A119 

\bibitem[{{Blaskiewicz} {et~al.}(1991){Blaskiewicz}, {Cordes}, \&
  {Wasserman}}]{BCW91}
{Blaskiewicz}, M., {Cordes}, J.~M., \& {Wasserman}, I. 1991, ApJ, 370, 643

\bibitem[{{Cordes}(1978)}]{Cordes78}
{Cordes}, J.~M. 1978, ApJ, 222, 1006

\bibitem[{{Deshpande} \& {Rankin}(1999)}]{DR99}
{Deshpande}, A.~A., \& {Rankin}, J.~M. 1999, ApJ, 524, 1008

\bibitem[{{Deshpande} \& {Rankin}(2001)}]{DR01}
---. 2001, MNRAS, 322, 438

\bibitem[{{Drake} \& {Craft}(1968)}]{DC68}
{Drake}, F.~D., \& {Craft}, H.~D. 1968, Nature, 220, 231

\bibitem[Lorimer, \& Kramer(2004)]{handbook04}
Lorimer, D.~R., \& Kramer, M.\ 2004, Handbook of pulsar astronomy


\bibitem[{{Dyks} {et~al.}(2004){Dyks}, {Rudak}, \& {Harding}}]{DRH04}
{Dyks}, J., {Rudak}, B., \& {Harding}, A.~K. 2004, ApJ, 607, 939

\bibitem[{{Edwards} \& {Stappers}(2002)}]{ES02}
{Edwards}, R.~T., \& {Stappers}, B.~W. 2002, A\&A, 393, 733

\bibitem[{{Edwards} \& {Stappers}(2003)}]{ES03}
---. 2003, A\&A, 410, 961

\bibitem[{{Edwards} {et~al.}(2003){Edwards}, {Stappers}, \& {van
  Leeuwen}}]{Edwards03}
{Edwards}, R.~T., {Stappers}, B.~W., \& {van Leeuwen}, A.~G.~J. 2003, A\&A,
  402, 321

\bibitem[{{Gajjar} {et~al.}(2014){Gajjar}, {Joshi}, {Kramer}, {Karuppusamy}, \&
  {Smits}}]{Gajjar14}
{Gajjar}, V., {Joshi}, B.~C., {Kramer}, M., {Karuppusamy}, R., \& {Smits}, R.
  2014, ApJ, 797, 18

\bibitem[{{Gangadhara} \& {Gupta}(2001)}]{GG01}
{Gangadhara}, R.~T., \& {Gupta}, Y. 2001, ApJ, 555, 31

\bibitem[\protect\citename{{Gupta} \& {Gangadhara}, }2003]{GG03}
{Gupta}, Y., \& {Gangadhara}, R.~T. 2003, ApJ, 584, 418

\bibitem[{{Gil} \& {Sendyk}(2000)}]{GS00}
{Gil}, J.~A., \& {Sendyk}, M. 2000, ApJ, 541, 351

\bibitem[{{Herfindal} \& {Rankin}(2007)}]{HR07}
{Herfindal}, J.~L., \& {Rankin}, J.~M. 2007, MNRAS, 380, 430

\bibitem[{{Herfindal} \& {Rankin}(2009)}]{HR09}
---. 2009, MNRAS, 393, 1391

\bibitem[{{Kijak} \& {Gil}(2003)}]{KG03}
{Kijak}, J., \& {Gil}, J. 2003, 397, 969

\bibitem[Maan(2018)]{Maan18}
Maan, Y.\ 2018, Pulsar Astrophysics the Next Fifty Years, 337, 366 

\bibitem[{{Maan} \& {Deshpande}(2014)}]{MD14}
{Maan}, Y., \& {Deshpande}, A.~A. 2014, ApJ, 792, 130

\bibitem[{{Maan} {et~al.}(2013){Maan}, {Deshpande}, {Chandrashekar},
  {Chennamangalam}, {Raghavendra Rao}, {Somashekar}, {Anderson}, {Ezhilarasi},
  {Sujatha}, {Kasturi}, {Sandhya}, {Bauserman}, {Duraichelvan}, {Amiri},
  {Aswathappa}, {Barve}, {Sarabagopalan}, {Ananda}, {Beaudet}, {Bloss},
  {Dhamnekar}, {Egan}, {Ford}, {Krishnamurthy}, {Mehta}, {Minter}, {Nagaraja},
  {Narayanaswamy}, {O'Neil}, {Raja}, {Sahasrabudhe}, {Shelton}, {Srivani},
  {Venugopal}, \& {Viswanathan}}]{Maan13}
{Maan}, Y., {Deshpande}, A.~A., {Chandrashekar}, V., {et~al.} 2013, ApJS, 204,
  12

\bibitem[{{Mitra} \& {Rankin}(2008)}]{MR08}
{Mitra}, D., \& {Rankin}, J.~M. 2008, MNRAS, 385, 606

\bibitem[{{Rankin}(1983)}]{Rankin83a}
{Rankin}, J.~M. 1983, ApJ, 274, 333

\bibitem[{{Rankin}(1993)}]{Rankin93a}
---. 1993, ApJ

\bibitem[{{Rankin} \& {Ramachandran}(2003)}]{RR03}
{Rankin}, J.~M., \& {Ramachandran}, R. 2003, ApJ, 590, 411

\bibitem[{{Rankin} {et~al.}(2003){Rankin}, {Suleymanova}, \&
  {Deshpande}}]{RSD03}
{Rankin}, J.~M., {Suleymanova}, S.~A., \& {Deshpande}, A.~A. 2003, MNRAS, 340,
  1076

\bibitem[{{Ruderman} \& {Sutherland}(1975)}]{RS75}
{Ruderman}, M.~A., \& {Sutherland}, P.~G. 1975, ApJ, 196, 51

\bibitem[{{Smith} {et~al.}(2013){Smith}, {Rankin}, \& {Mitra}}]{SRM13}
{Smith}, E., {Rankin}, J., \& {Mitra}, D. 2013, MNRAS, 435, 1984

\bibitem[{{Srostlik} \& {Rankin}(2005)}]{SR05}
{Srostlik}, Z., \& {Rankin}, J.~M. 2005, MNRAS, 362, 1121

\bibitem[{{Szary} \& {van Leeuwen}(2017)}]{Szary17}
{Szary}, A., \& {van Leeuwen}, J. 2017, ApJ, 845, 95

\bibitem[{{van Leeuwen} \& {Timokhin}(2012)}]{JvL12}
{van Leeuwen}, J., \& {Timokhin}, A.~N. 2012, ApJ, 752, 155

\end{thebibliography}
\end{document}